\documentclass[12pt]{article}

\advance \textheight by \topskip
\makeatletter
\@addtoreset{equation}{section}
 
\makeatother

\def\R{\mathbb R}

\def\J{\mathfrak J}

\usepackage{amsmath,amssymb}
\begin{document}
\title{
%\begin{flushright}
%{\small USACH-FM-03/02}\\[1.0cm]
%\end{flushright}
{\bf Conformal Generators and Doubly Special Relativity Theories}}
\author{{\sf Carlos Leiva}\thanks{
E-mail: cleivas@uta.cl}
\\
{\small {\it Departamento de F\'{\i}sica, Universidad de
Tarapac\'{a}, Casilla 7-D Arica, Chile}}}

\date{}

%%*********************** TITLE **********************

%%%%%%%%%%%%%%%%%%%%%%%%%%%%%%%%%%%%%%%%%%%%%%%%%%%%%%%%%%%%
%%%%%%%%%%%%%%%%%%%%%%%%%%%

%%*********************** AUTHORS **********************

%%%%%%%%%%%%%%%%%%%%%%%%%%%%%%%%%%%%%%%%%%%%%%%%%%%%%%%%%%%%
%%%%%%

%%%%%%%%%%%%%%%%%%%%%%%%%%%%%%%%%%%%%%%%%%%%%%%%%%%%%%%%%%%%
%%%%%%

\maketitle

%********************** ABSTRACT ***********************

\begin{abstract}
In this paper,  the relation between the modified Lorenz boosts,
proposed in the doubly relativity theories and a linear
combination of  Conformal Group generators in $R^{1,d-1}$ is
investigated. The introduction of a new generator is proposed in
order  to deform the Conformal Group to achieve the connection
conjectured. The new generator is obtained trough a formal
dimensional reduction from a free massless particle  living in a
$R^{2,d}$ space. Due this treatment it is possible to say that
even DSR theories modify light cone structure in $R^{1,d-1}$, it
 could remains, in some cases, untouched in $R^{2,d}$.
\end{abstract}

%\newpage
%********************* SECTION 1 ***********************
\section{Introduction}
There are some evidences that the special relativity must be
modified. It has been claimed that Lorentz symmetry breaking could
be observable in the future in high energy cosmic ray spectra
\cite{giovanni}, and  corrections in the dispersion relation
$E^2-p^2=m^2$ have been proposed \cite{piran}. On the other hand,
quantum gravity models  suggest that it could be desirable to
review the Lorentz invariance relations. Indeed there are special
features in these theories: the Planck longitude $l_p=\sqrt{\hbar
G/c^3}$, its associated time scale $t_p=l_p/c$ and the Planck
energy $E_p=\hbar /t_p$. They are thresholds beyond the physics
should change dramatically. However, absolute values of longitude,
time or energy are not in agreement with the Lorentz
transformations, because  they  are not the same  for different
observers in different frames. Of course, whatever the deformation
of the Lorentz boost introduced be, it must fit smoothly with the
usual experience at energies  far from Planck scales.

Several solutions for these problems have been proposed, in
particular doubly special relativity (DSR) theories
\cite{mag},\cite{giovanni2},\cite{kow}. They are generalizations
of the special relativity with two observer independent scales.
Starting  with the Fock-Lorentz transformations in the position
space \cite{fock}, \cite{manida}, a modification of Lorentz boosts
in the momentum space was proposed by Magueijo and Smolin
\cite{mag}, with a very similar approach. The treatment of the
problem in the momentum space has been preferred because of the
introduction of deformed dispersion relations, however when this
choice is done, recovering the position space dynamics can be a
highly problematic task due to the loss of linearity. Recently
Kimberly, Magueijo and Medeiros, \cite{mag2}, have shown methods
 for obtaining position space of
non linear relativity models from the usual momentum space
formulation, using a free field theory.

Nowadays, DSR theories are of increased interest because they can
be useful as a new tool in gravity theories, in Cosmology  as an
alternative to inflation, \cite{mof}, \cite{alb}, and in other
fields like propagation of light \cite{ku}, that is related, for
instance, to cosmic microwave background radiation.

In this paper it is demonstrated that it is possible, formally, to
obtain Fock-Lorentz and Magueijo-Smolin deformations, trough a
reduction process and using the conformal group generators as the
generators of the deformed Lorentz algebra.

More specifically, it is conjectured that the deformations of the
Lorentz algebra in the Fock-Lorentz formulation,  can be treated
as a transformation made by a linear combination of conformal
group generators and the momentum case, can be understood as the
same process, but the inclusion of a new generator is needed. This
new generator can be constructed from the same theory, but some
new aspects of the dimensional reduction process proposed for the
formers must be added. Finally the nonlinear induced algebra is
shown. So, a relationship between space and momentum treatments is
worth investigating as  becoming from a bigger, underlaying theory
that could bear the whole features of DSR theories.

The paper is organized as follows. In section 2, the conformal
symmetry group of a $d$ space, massless relativistic particle is
shortly reviewed. In section 3, isometries of a $d+2$ space plus a
dimensional reduction process is recalled as underlying in the
origin of those symmetries. In section 3, the position space DSR
transformation are presented as combination of the conformal
generators. Then, the momentum space DSR transformations are
obtained using an extra generator. In section 4, the origin of
this new generator is conjectured and the algebra including this
new generator is shown. Conclusions and comments are presented in
section 5.

%%%%%%%%%%%%%%
%%%%%%%%%%%%%%%%%%
%%%%%%%%%%%%%%

\section{\protect\bigskip Conformal Symmetry of Massless
Particle}
%\bigskip\cdot

The infinitesimal transformations
\begin{equation}
\delta x^\mu = \omega^\mu{}_\nu x^\nu  + \alpha^\mu + \beta x^\mu
+ 2(x\gamma)x^\mu - x^2\gamma^\mu \label{inf1}
\end{equation}
generate the conformal symmetry, $ds^2\rightarrow
ds'{}^2=e^{2\sigma}ds^2$, on the $d$-dimensional Minkowski space
$\R^{1,d-1}$ with metric
$$
ds^2=dx^\mu dx^\nu\eta_{\mu\nu}=-dx_0^2+\sum_{i=1}^{d-1}dx^{ 2}_i.
$$
Here the parameters $\omega^\mu{}_\nu$, $\alpha^\mu$, $\beta$ and
$\gamma^\mu$ correspond to the Lorentz rotations, space-time
translations, scale and special conformal transformations. Due to
a nonlinear (quadratic) in $x^\mu$ nature of the two  last terms
in (\ref{inf1}), the finite version of the special conformal
transformations,
\begin{equation}
x'{}^\mu=\frac{x^\mu-\alpha^\mu x^2}{1 -2\alpha x + \alpha^2 x^2},
\label{confin}
\end{equation}
is not defined globally, and to be well defined requires a
compactification of $\R^{1,d-1}$ by including the points at
infinity.

On the classical phase space with canonical Poisson bracket
relations $\{x_\mu,$ $p_\nu\}=\eta_{\mu\nu}$,
$\{x_\mu,x_\nu\}=\{p_\mu,p_\nu\}=0$, the transformations
(\ref{inf1}) are generated by
\begin{equation}
M_{\mu \nu } = x_{\mu }p_{\nu }-x_{\nu }p_{ \mu }, \qquad P_{\mu }
= p_{\mu },\qquad D = x^{\mu }p_{\mu }, \qquad K_{\mu } = 2x_{\mu
}(xp)-x^{2}p_{\mu }. \label{genconf}
\end{equation}
The generators (\ref{genconf}) form the conformal algebra
\begin{eqnarray}
&\{M_{\mu\nu},M_{\sigma\lambda}\}= \eta_{\mu\sigma}M_{\nu\lambda}-
\eta_{\nu\sigma}M_{\mu\lambda}+ \eta_{\mu\lambda}M_{\sigma\nu}-
\eta_{\nu\lambda}M_{\sigma\mu},&\notag\\
&\{M_{\mu \nu },P_{\lambda }\} = \eta_{\mu \lambda }P_{ \nu }-
\eta_{\nu\lambda }P_{\mu },\qquad \{M_{\mu \nu },K_{\lambda }\} =
\eta_{\mu \lambda }K_{ \nu }-
\eta_{\nu\lambda }K_{\mu },&\notag\\
&\{D, P_{\mu }\} =P_{\mu },\qquad
\{D, K_{\mu }\} =-K_{\mu },&\notag\\
&\{ K_{\mu },P_{\nu }\} = 2(\eta_{\mu \nu }D+M_{\mu \nu}),&
\notag \\
&\{D, M_{\mu\nu}\} = \{P_\mu,P_\nu\}= \{ K_{\mu },K_{\nu }\} =0.&
\label{confalg}
\end{eqnarray}
The algebra (\ref{confalg}) is isomorphic to the algebra
$so(2,d)$, and by defining
\begin{equation}
J_{\mu \nu }=M_{\mu \nu },\quad
 J_{\mu d}=\frac{1}{2}(P_\mu + K_\mu),\quad
 J_{\mu (d+1)}=\frac{1}{2}(P_\mu-K_\mu),
 \quad
 J_{d(d+1)}=D,
 \label{ident}
\end{equation}
can be put in the standard form
\begin{equation}
\{J_{AB},J_{LN}\}= \eta_{AL}J_{BN}- \eta_{BL}J_{AN}+
\eta_{AN}J_{LB}- \eta_{BN}J_{LA} \label{so24}
\end{equation}
with $A,B=0,1,d,d+1$, and
\begin{equation}
\eta_{AB}=diag (-1,+1,\ldots,+1,-1). \label{etaab}
\end{equation}

The phase space constraint $ \varphi_m\equiv p^2+m^2=0 $
describing the free relativistic particle of mass $m$ in
$\R^{1,d-1}$ is invariant under the Poincar\'e transformations,
$\{M_{\mu\nu},\varphi_m\}= \{P_\mu,\varphi_m\}=0$. Unlike the
$P_\mu$ and $M_{\mu\nu}$, the generators of the scale and special
conformal transformations commute weakly with $\varphi_m$ only in
the massless case $m=0$: $\{D,\varphi_0\}=2\varphi_0=0$,
$\{K_\mu,\varphi_0\}=4x_\mu\varphi_0=0$.

\section{Dimensional Reduction}

 It was demonstrated in \cite{LP}, that a dimensional reduction process from
a massless particle living in $(d+2)$ dimensional space
$\R^{2,d}$,  with  coordinates $H^A$ and metric (\ref{etaab}), and
introducing  canonical momenta $\Pi_A$,
($\{H_A,\Pi_B\}=\eta_{AB}$) can transform  the $so(2,d)$
generators
\begin{equation}
\J_{AB}=H_A \Pi_B - H_B \Pi_A. \label{jxp},
\end{equation}
into the conformal generators in $(d)$-dimensional space
$\R^{1,d-1}$. In order to achieve this, three constrains are added
to the theory:

\begin{eqnarray}
&\phi_0\equiv \Pi_A\Pi^A=0,&
\label{xp2}\\
&\phi_1\equiv H^{A} H_{A}=0,\qquad \phi_2\equiv
H^{A}\Pi_A=0.&\label{xp1}
\end{eqnarray}

The process is implemented trough suitable canonical
transformations and then a dimensional reduction is performed
 onto  the surface defined by $\phi_1$
and $\phi_2$. So in terms of new $\R^{2,d}$ canonical variables
$\tilde H^\mu$, $\tilde\Pi_\mu$ with  $\{\tilde H^\mu,
\tilde\Pi_\mu\}=1$ as defined in \cite{LP}, the $so(2,d)$
generators are the following:

\begin{eqnarray}
\J_{\mu\nu}=\tilde H_\mu\tilde\Pi_\nu- \tilde H
_\nu\tilde\Pi_\mu,\quad \J_{\mu +}=\tilde\Pi_\mu,\quad
\J_{d(d+1)}=\tilde H_\mu\tilde\Pi^\mu+
2\frac{\tilde\Pi_+}{\tilde H^-}\phi_1 -\phi_2, \\
\J_{\mu -}=2(\tilde H_\nu\tilde\Pi^\nu)\tilde H_\mu
-(\tilde\Pi_\nu\tilde H^\nu)\tilde\Pi_\mu +\frac{1}{\tilde H
^{-2}}(\tilde\Pi_\mu +4\tilde\Pi_+\tilde H^-\tilde H_\mu)\phi_1
-2\tilde H_\mu\phi_2.\notag \label{jotas}
\end{eqnarray}

To execute  this process it is supposed that $\tilde H^- \neq 0$.
The constraints now, look like:
\begin{eqnarray}
&\tilde\Pi_\mu\tilde\Pi^\mu=0,&\label{tp2}
\\
&\tilde H^+=0,\qquad \tilde\Pi_-=0.&\label{x+p-}
\end{eqnarray}

The variables $\tilde\Pi_\mu$ and $\tilde H_\mu$ are observable
(gauge invariant) variables with respect to the constraints
(\ref{tp2}) and (\ref{x+p-}), whereas $\tilde H^-= H^-$ and
$\tilde\Pi_+=\Pi_+$ are not. They can be removed by introducing
the constraints
\begin{equation}
\phi_3\equiv H^-+1=0,\quad \phi_4\equiv\Pi_+=0 \label{gaugex}
\end{equation}
as gauge fixings of (\ref{x+p-}) and the reduction to the physical
surface achieves the conformal generators in $\R^{1,d-1}$. Now,
with the identification $x^\mu=\tilde H^\mu$ and
$p_\mu=\tilde\Pi_\mu$ and (\ref{tp2}),  the original massless
relativistic particle is retrieved.

\section{DSR transformations}
\subsection{Fock-Lorentz transformations}
The Fock-Lorentz transformation \cite{fock}, \cite{manida}, are
introduced as general linear-fractional transformations between
spatial and temporal coordinates and the result is leaded by
symmetry arguments to the final form:

\begin{eqnarray}
t'= \frac{\gamma (u)
(t-\mathbf{u}\mathbf{r}/c^2)}{1-(\gamma(u)-1)ct/R
+\gamma(u)\mathbf{u}\mathbf{r}/Rc},\\
\mathbf{r}\|=\frac{\gamma (u) (\mathbf{r}\|
-\mathbf{u}t)}{1-(\gamma(u)-1)ct/R
+\gamma(u)\mathbf{u}\mathbf{r}/Rc},\\
\mathbf{r}\bot=\frac{\mathbf{r}\bot}{1-(\gamma(u)-1)ct/R
+\gamma(u)\mathbf{u}\mathbf{r}/Rc},\\
\end{eqnarray}
where $\mathbf{r}\|$ is the component of the position vector
parallel to the boost $\mathbf{u}$ and $\mathbf{r}\bot$ is the
perpendicular one. Where  $\gamma$  is the Lorentz factor
$\sqrt{1-u^2/c^2}$ and $R$ is a constant with the dimension of
length.

These transformations can be seen as Lorentz transformations for
the quantities:

\begin{equation}
\tilde{t}= \frac{t}{1+ct/R},\qquad
\mathbf{\tilde{r}}=\frac{\mathbf{r}}{1+ct/R} \label{trtilde}
\end{equation}

They can be treated as results of a infinitesimal transformations:

\begin{equation}
\delta x^\mu= \alpha^\nu \{x^\mu ,2x_\nu (xp)\} \label{trapos}
\end{equation}
where $x^0=ct$.

In order to obtain this generator, it can be represented as the
projection of
\begin{equation}
(\J_{\mu -} + \J_{\mu +}\J_{d (d+1)}), \label{jnolin}
\end{equation}
 in $R^{d+2}$ space, wich is
a nonlinear combination of the isometries there. However $2x_\nu
(xp)$ is not a symmetry of the Klein-Gordon equation because
$\{2x_\nu (xp), \varphi_0\}=4p_\mu (xp) + 4x_\mu \varphi_0 \neq
0$. So, it is not possible  to include it into  the generator set
of the massless relativistic particle symmetries.

An alternative, an linear solution is to see (\ref{trtilde}) as
  (\ref{confin}) reduced to  the surface
$x^2=0$ and with the choice $\alpha=(-1/2R,0,0,0)$ and identifying
$x^0$ with $ct$. In this sense, the Fock Lorentz transformations
can be seen as a linear combination of conformal generators $$
M_{0 \mu } + K_{ \mu },$$
 that is a projection of
$$(\J_{\mu \nu} + \J_{\mu -})$$
which is linear too, for a particle living on the cone
$x_{1}^{2}+x_2^2+x_3^2=x_0^2$.

\subsection{Momentum space transformations}

 On the other hand, the Lorentz boost proposed in \cite{mag}

\begin{equation}
K^i= L_{0 i} +l_{p} p^{i} p_{\mu} \frac{\partial} { \partial
p_{\mu}} \label{lomod}
\end{equation}
where $l_p$ is the Planck length,  can be exponentiated  as

\begin{equation}
K^i= U^{-1}(p_0)L_{0 i}U(p_0)
\end{equation}
where $U(p_0)=exp(l_p p_0 p_{\mu} \frac{\partial}{ \partial
p_{\mu}})$, and the action of $U(p_0)$ over $p_\mu$ is

\begin{equation}
U(p_0)p_\mu= \frac{p_\mu}{1-l_p p_0} \label{p0}
\end{equation}

The generator added to the Lorentz boost in (\ref{lomod}) is the
operational representation of $p^{i}(xp)$ and it can be seen as
part of some $\tilde{K}_\mu = 2p_\mu (xp)-p^2 x_\mu$ generator
that, with the constrain $\varphi_0$ in mind, produces the
following transformation :

\begin{equation}
p'{}^\mu=-\frac{p^\mu-\alpha^\mu \varphi_0}{1 -2\alpha p +
\alpha^2 \varphi_0}, \label{confinp}
\end{equation}
to achieve the identification with (\ref{p0}), the transformation
parameter must be $\alpha^\mu= (-l_p/2,0,0,0)$.

It is possible then, to prescribe  how to obtain the position and
momenta transformation using conformal generators with some
$\tilde{K}$ added. Where does this new generator come from? Some
clues will be seen in the next section.

%\newpage

\section{Extended Non-linear Conformal Algebra}

It is possible, trough a rather tricky process, to introduce the
new generator $\tilde{K}_{\mu }$. To do this, we can perform a new
canonical transformation in (\ref{jotas}), $\tilde{H}^{A}
\rightarrow -\breve{\Pi}^A$ and $\tilde{\Pi}^{A} \rightarrow
\breve{H}^A$ and now to identify $x^\mu=\breve H_\mu$ and
$p_\mu=\breve\Pi_\mu$ . Doing this, $\J_{\mu -}$ is projected as
$\tilde{K}$, and  $\J_{\mu +}$ is projected as $x^\mu$, but the
particle  has now the constrain $x^2=0$.

If we want to use this generator on the $\varphi_0=p^2=0$ surface
and we wish to maintain this kind of description as underlying in
the origin of the generator, it is necessary to set $xp\neq 0$,
otherwise $D$, $K$ and $\tilde{K}$ become zero and the theory
becomes trivial. This condition means that $x^2=0$ can be seen as
a gauge fixing for $p^2=0$, because $\{x^2,p^2\}=4xp\neq 0$.

 $\tilde{K}$ is a symmetry of the Klein-Gordon equation too, in
fact: $\{\tilde{K},\varphi_0\}=2\varphi_0 P_\mu=0$. So the
relation of $\tilde{K}$ with the other generators is.

 After the inclusion of $\tilde{K}$, aside the
relations (\ref{confalg}) we have:

\begin{eqnarray}
\{M_{\mu \nu },\tilde{K}_{\lambda }\} &=& \eta_{\mu \lambda }\tilde{K}_{ \nu}- \eta_{\nu \lambda }\tilde{K}_{\mu },\notag\\
 \{D, \tilde{K}_{\mu }\} &=&\tilde{K}_{\mu }, \notag\\
\{ \tilde{K}_{\mu },P_{\nu }\} &=& 2P_\mu P_\nu- \varphi_0 \eta_{\mu \nu},\notag \\
\{ \tilde{K}_{\mu },\tilde{K}_{\nu }\} &=&0, \notag \\
\{ \tilde{K}_{\mu },K_{\nu }\} &=& 4(M_{\mu \nu}-\eta_{\mu
\nu})D-2P_\mu K_\nu -\varphi_0 (q^2\eta_{\mu \nu}-2q_\mu q_\nu)
\label{confalgdef}
\end{eqnarray}

These relations generate a non-linear algebra on the constrain
$\varphi_0$ surface

\section{Discussion and outlook}

To conclude, let us summarize the obtained results
and discuss shortly some problems that deserve a further
attention.

Transformations of DSR theories can be partially achieved by a
suitable combination of conformal generators. To accomplish this
in the momentum space, an extra generator must be added to the
conformal group. This one can be obtained, formally, from the same
$R^{d+2}$ massless particle theory, as the others, but a rather
tricky process must be performed. Anyway, this generator is a
symmetry of the Klein-Gordon equation and  when it is included, a
non-linear algebra arise just on mass shell.

An important feature of this treatment is the evidence that even
the Lorentz invariance is broken in $R^d$ by DSR  theories, when
we are dealing with some special cases as a massless relativistic
particle living in the cone $x^2=0$, it could remains untouched in
$R^{d+2}$ because the deformed Lorentz boost in the former are
linear combinations of the images of the isometries of the latter.
Even although  we can accept this  idea, there are not clues about
the conditions for the parameters, in order to include $l_p$, this
must be a future task. On the other hand, it will be important to
investigate the geometry of the systems here detailed and to
analyze the case when totally arbitrary parameters are included.

Of course, if a discrete space-time is claimed, the notion of a
point particle must be abandoned, so all classic treatments must
be replaced by a suitable theory. In order to do this, the ideas
proposed here must be translated to the field theories ambit.

\vskip 0.5cm {\bf Acknowledgements} \vskip 5mm I want to thank to
the Departamento de F\'{\i}sica de la Universidad de Tarapac\'{a}
and to N. Cruz for calling my attention to DSR theories.


\begin{thebibliography}{99}



\bibitem{giovanni}
G. Amelino-Camelia and Tsvi Piran, \textit{Planck-scale
deformation of Lorentz symmetry as a solution to the UHECR and
$Tev-\gamma$ paradoxes},  Phys. Rev.D64(2001) 036005
[astro-ph/0008107].

\bibitem{piran}
G. Amelino-Camelia and T. Piran, Phys. Rev. D64 (2001) 036005.

\bibitem{giovanni2}
G. Amelino-Camelia, Int. J. Mod. Phys. D11, 35, 2002,
[gr-qc/0012051], Ohys. Lett. B510, 255-263, 2001.

\bibitem{kow}
J. Kowalski-Glikman, Phys. Lett. A286, 391-394, 2001,
[hep-th/0102098]; N.R. Bruno, G. Amelino-Camelia, J.
Kowalski-Glikman, Phys. Lett. B522, 133-138,2001.

\bibitem{mof}
J. Moffat, Int. J. of Physics D \textbf{2}, 351 (1993); J. Moffat,
Foundations of Physics, \textbf{23}, 411 (1993)


\bibitem{alb}
A. Albrecht and J. Magueijo, Phys. Rev. D \textbf{59}, 043516
(1999)

\bibitem{fock}
V. Fock, \textit{The theory of space-time and gravitation},
Pergamon Press,1964.

\bibitem{manida}
S.N. Manida, gr-qc/9905046

\bibitem{mag}
J. Magueijo and L. Smolin, Phys. Rev. Lett. \textbf{88},
190403(2002) [hep-th/0112090]

\bibitem{mag2}
D. Kimberly, J. Magueijo and J. Medeiros [gr-qc/0303067]

\bibitem{ku}
Sung Ku Kim, Sun Myong Kim, Chaiho Rim and Jae Hyung Yee,
\textit{Propagation of Light in Doubly Special Relativity}
[gr-qc/0401078]

\bibitem{LP}
C.~Leiva and M.~S.~Plyushchay, \textit{Conformal Symmetry of
Relativistic and Nonrelativistic Systems and AdS/CFT
Correspondence}, Annals of Physics 307 0310 (2003) 372-391.
[arXiv:hep-th/0301244].
%%CITATION = HEP-TH 0304257;%%




\end{thebibliography}
\end{document}